\documentclass[preprint,12pt]{elsarticle}

\usepackage{amssymb}
\usepackage[utf8]{inputenc}
\usepackage[english]{babel}
\usepackage{pdflscape}
\usepackage{listings}
\usepackage{scrextend}
\usepackage{multirow}
\usepackage{colortbl}
\usepackage[table]{xcolor}
\usepackage{longtable}
\usepackage{verbatim}

\journal{Journal of Systems and Software}

\begin{document}

\begin{frontmatter}



\title{Secondary Studies in the Academic Context: A Systematic Mapping and Survey}

 \author{Katia Romero Felizardo, Érica Ferreira de Souza, Bianca Minetto Napole\~ao}
 \address{Department of Computer, Federal University of Technology - Paraná - UTFPR,  Brazil\\ Cornélio Procópio - Paraná}
 \ead{ {biancanapoleao@alunos.utfpr.edu.br, {ericasouza, katiascannavino}@utfpr.edu.br}}

 \author{Nandamudi Lankalapalli Vijaykumar}
 \address{National Institute for Space Research (INPE) - Laboratory of Computing and Applied mathematics (LAC) - \\ São José dos Campos - São Paulo and Federal University of São Paulo (Unifesp) - Institute of Science \& Technology (ICT) - \\São José dos Campos - São Paulo}
 \ead{vijay.nl@inpe.br, vijaykumar@unifesp.br}

 \author{Maria Teresa Baldassarre}
 \address{Universit\`a degli Studi di Bari Aldo Moro/Universit\`a di Bari\\ Piazza Umberto I -- 70121, P.I.01086760723/C.F.80002170720\\ Bari/Italy}
 \ead{mariateresa.baldassarre@uniba.it}

\begin{abstract}
	\textbf{Context:} Several researchers have reported their experiences in applying secondary studies (Systematic Literature Reviews - SLRs and Systematic Mappings - SMs) in Software Engineering (SE). However, there is still a lack of studies discussing the value of performing secondary studies in an academic context.
	\textbf{Goal:} The main goal of this study is to provide an overview on the use of secondary studies in an academic context.
	\textbf{Method:} Two empirical research methods were used. Initially, we conducted an SM to identify the available and relevant studies on the use of secondary studies as a research methodology for conducting SE research projects. Secondly, a survey was performed with 64 SE researchers to identify their perception related to the value of performing secondary studies to support their research projects.
	\textbf{Results:} Our results show benefits of using secondary studies in the academic context, such as, providing an overview of the literature as well as identifying relevant research literature on a research area enabling to find reasons to explain why a research project should be approved for a grant and/or supporting decisions made in a research project. Difficulties faced by SE graduate students with secondary studies are that they tend to be conducted by a team and it demands more effort than a traditional review.
	\textbf{Conclusions:} Secondary studies are valuable to graduate students. They should consider conducting a secondary study for their research project due to the benefits and contributions provided to develop the overall project. However, the advice of an experienced supervisor is essential to avoid bias. In addition, the acquisition of skills can increase student's motivation to pursue their research projects and prepare them for both academic or industrial careers. 
\end{abstract}


\end{frontmatter}

\section{Introduction} \label{introduction}

Evidence-Based Software Engineering (EBSE) refers to the adoption of appropriate research methods to build a body of knowledge about when, how, and in what context methods or tools are more appropriate to be used for practicing Software Engineering (SE). EBSE was first introduced in 2004 as a means to advance and improve the discipline of SE \cite{Kitchenham15}. In this context, secondary studies, such as, Systematic Review (SR) (a.k.a Systematic Literature Review (SLR)) and Systematic Mapping Study (SMS) have provided mechanisms to identify and aggregate research evidence \cite{Kitchenham15}. While SLR has been used to provide a complete and fair evaluation of the state of evidence related to a specific topic of interest, SM (also known as a scoping review) is a more open form of SLR, providing an overview of a research area to assess the quantity of existing evidence on a topic of interest. 

Since their introduction in the SE field, SLRs and SMSs have been gaining importance \cite{Dyba05}. Their use is increasing as a method for conducting secondary studies in SE \cite{Silva10, Zhang11a}. In other fields, secondary studies have also gained wide acceptance. For example, in social sciences, medicine and related professions, including biochemistry, many graduate students adopt secondary studies as a basis in their dissertations or theses  \cite{Daigneault14,Puljak17}. 

In SE, normally, students at both undergraduate and graduate levels are required to conduct traditional literature reviews, which, as they are frequently seen in the literature, do not use a systematic approach; hence one cannot rule out that the choice of studies and the conclusions drawn could be biased, thus providing readers with a distorted view about the state of knowledge regarding the area at the focus of the review. On the contrary, a secondary study uses a systematic process to identify, assess and interpret all available research evidence to provide reliable answers to a particular research question. 

Over the past decade, most research in EBSE has emphasized potential advantages and disadvantages of the use of secondary studies \cite{Riaz10, Lavallee14, Kuhrmann17} and presented challenges and lessons learned \cite{Kitchenham13, Imtiaz13}. Several studies have attested that a secondary study is a valuable research mechanism for providing knowledge of a given topic and supporting the identification of gaps for future research \cite{Dyba05, Kitchenham11, Zhang11a}. However, what is not clear yet is how this knowledge supports conducting M.Sc./Ph.D. research projects. There were no actual data that would indicate how SE graduate (M.Sc. or Ph.D.) students have used a secondary study as research methodology for their dissertation/thesis and how the student and his/her supervisor (or any other researcher) perceives it. This article also addresses this gap by offering an overview on the use of secondary studies for developing M.Sc./Ph.D research. 

This article reflects upon the use of secondary studies in developing academic projects. Therefore, the purpose of this study is to explore SE researchers perceptions, in particular M.Sc./Ph.D. students and their supervisors, about the value of secondary studies and how these perceptions impact decisions on conducting their research. 

We believe that understanding the benefits and challenges of using secondary studies in an academic context is useful for both students who are starting their research projects, and those that will not continue their academic career (future practitioners), since they can all base their decisions on the best available evidence provided by secondary studies. Specifically, the main goals of this research are to:

\begin{enumerate}		
	\item evaluate the value of secondary studies in an academic context; 
	\item reinforce the importance of a secondary study in conducting a research project;
	\item discuss how findings of a secondary study are used for conducting a research project; 
	\item report experiences of M.Sc./Ph.D. students conducting a secondary study as part of their research project; and
	\item inspire graduate students to use EBSE, in particular, SLR/SMS in their research projects.	
\end{enumerate}

Based on these goals, some of the questions we intend to clarify as well as the rationale for considering them are presented in Table \ref{table:RQ}.

\begin{table*}[!ht]
	\caption{Research questions and their rationales} \label{table:RQ}
	\vskip 0.3cm
	\centering
	\begin{tabular}{p{0.8cm}||p{4cm}|p{7.5cm}}
		\hline
		\textbf{N$^{o}$} & \textbf{Research question} & \textbf{Rationale} \\ \hline \hline
		
		\textbf{RQ1} &  When and where the studies on the academic application of secondary studies in SE have been published? &  This research question provides an understanding on whether there are specific publication sources for these studies, and when they have been published. \\ 
		\hline
		\textbf{RQ2} & What is the academic purpose in applying secondary studies in SE? & This question looks for the purposes declared in the studies for applying secondary studies in an academic context. This is important to point out why such studies have been accomplished. \\
		\hline		
		\textbf{RQ3} & What are the benefits reported by SLR/SMS authors related to the use of secondary studies in an academic context? & Provides an overview of the main benefits reported by authors related to the main advantages in applying secondary studies in an academic context. \\ 
		\hline
		\textbf{RQ4} & What are the problems reported by authors related to the use of secondary studies in an academic context?  &  Provides an overview of the main limitations reported by authors with respect to the main difficulties in research applying secondary studies in an academic context. \\ 
		\hline       
	\end{tabular}
\end{table*}

This work combines two empirical research methods. Firstly, we conducted an SMS on the use of secondary studies as a research instrument in SE. Secondly, for the identification of M.Sc./Ph.D. students perceptions, we applied questionnaire based surveys. The use of two methods enabled us to minimize the potential limitations of applying single research method to achieve our goals.

It was concluded that using secondary studies can help researchers: improve the rigor and breadth of literature reviews; demonstrate gaps in the literature;  and turn into a publishable research paper. Conducting a secondary study involves a number of practical challenges, including: extended period of time to perform the study; data extraction; definition of the research questions; lacking of domain and secondary studies' process knowledge and the development of the study protocol. 

The remainder of the paper is organized as follows. Section \ref{relatedWorks} presents the research method applied to perform the SMS; results are also presented. Section \ref{survey} presents the survey and its results, focusing on students' perceptions on the use of EBSE in their research project. Section \ref{discussions} discuss our main results and limitations of this work. Finally, Section \ref{conclusions} presents our concluding remarks.

\section{The use of secondary studies as a research instrument in SE: an SMS} \label{relatedWorks}

\begin{figure*}[ht]
	\centering
	\includegraphics [width=1\textwidth]{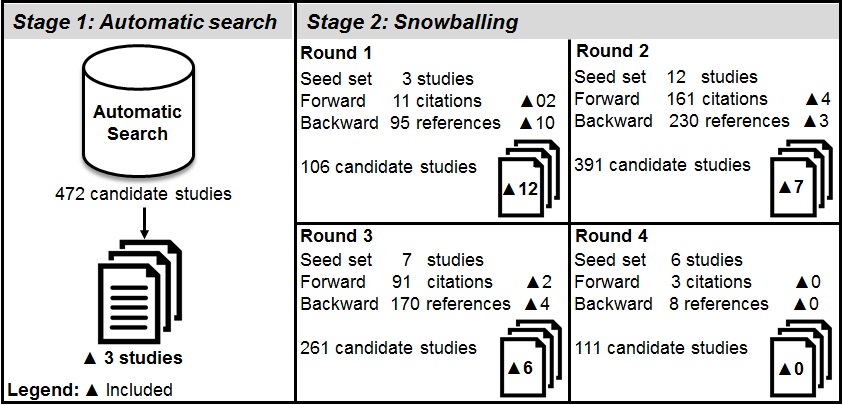}
	\caption{Searching for studies on the use of secondary studies as a research instrument in SE}
	\label{RelatedWork}
\end{figure*}

We conducted a systematic two-stage search, as shown in Figure \ref{RelatedWork}, in order to identify the current literature on the use of secondary studies as a research instrument in SE. 

First of all, in Stage 1, we created our search string, presented in Table \ref{table:string}. It was applied in \textit{Scopus}, one of the most commonly used databases in Computer Science \cite{Maplesden15} and \cite{Zhang11} with more than 60 million records. Scopus database includes papers of many international publishers, including Cambridge University Press, Institute of Electrical and Electronics Engineers (IEEE), Nature Publishing Group, Springer, Wiley-Blackwell and Elsevier, just to name a few. We executed the search string in three metadata fields: title, abstract and keywords. 

\begin{table*}[!ht]
	\caption{Keywords of the Search String -- Searching for related works} \label{table:string}
	\centering
	\begin{tabular}{c||p{10cm}}
		\hline
		\textit{\textbf{Search String}} & 
		(``evidence-based software engineering'' \textbf{OR} 
		``evidence based software engineering''  \textbf{OR} 
		``empirical software engineering''       \textbf{OR} 
		``software engineering literature''      \textbf{OR} 
		``systematic review''                    \textbf{OR} 
		``systematic literature review''         \textbf{OR}  
		``mapping study''                        \textbf{OR}  
		``mapping studies'') 
		\textbf{AND}
		student                                  \textbf{OR}
		novice                                   \textbf{OR}
		university                               \textbf{OR} 
		under-graduate                           \textbf{OR} 
		undergraduate                            \textbf{OR} 
		master	
		\\
		\hline
	\end{tabular}
\end{table*}

The selection criteria are organized in one Inclusion Criterion (IC) and seven Exclusion Criteria (EC):

\begin{itemize}
	\item[]\textbf{(IC1)} The study must be within the SE and discuss the use of secondary studies in an academic context.
\end{itemize} 

\begin{itemize}
	\item[] \textbf{(EC1)} The study does not have an abstract;\\  \textbf{(EC2)} The study is just published as an abstract;\\ 
	\textbf{(EC3)} The study is not written in English;\\
	\textbf{(EC4)} The study is an older version of other study already considered;\\ 
	\textbf{(EC5)} The study is not a primary study, such as editorials, summaries of keynotes, workshops, and tutorials;\\
	\textbf{(EC6)} The study is not in the scope of SE, although it uses secondary studies in an academic context; and\\ 
	\textbf{(EC7)} The study is in SE context, however, does not use secondary studies in an academic context.
\end{itemize} 

As illustrated in Figure \ref{RelatedWork} a total of 472 publications were returned. We considered the studies published until October 2018. The selection criteria (inclusion and exclusion criteria) were applied by the first author for title, abstract and keywords, leading to 154 studies. In the sequence, the selection criteria were applied jointly by two authors (first and second authors -- 100\% of agreement) considering the full text, resulting in a set of three studies from this stage.

During Stage 2, we used the three relevant studies selected for inclusion in Stage 1 as our seed set (starting set) for performing four iterations in snowballing where each iteration contained forward and backward snowballing. The citations (forward snowballing) were extracted with the help of search engines, such as \textit{Google Scholar}, \textit{ACM Digital Library} and \textit{IEEE Xplore Digital Library}. The results from iterations 1--4 of the snowballing approach application are shown in Figure \ref{RelatedWork} and detailed next.

Figure \ref{RelatedWork} shows that during the first round the three studies that formed the seed set were cited by 11 studies -- candidates for inclusion. Two of these 11 citations were included, when applying the same inclusion criteria used during Stage 1. The three articles totaled 95 references, out of these, ten were included. Round 2 started with 12 studies previously selected in round 1. They have been cited by a large number of studies (161). Out of these, four were included. 230 studies composed the references, however only three relevant ones were identified from these candidates. 91 studies have cited the seven studies previously included (round 2) and two were identified as relevant during round 3. Out of the 170 references, three were included. Thus, six (2: forward + 4: backward) studies were included. Finally, during round 4 no new studies were revealed. Therefore, the four iterations of the forward/backward snowballing approach identified 25 relevant studies. 

By combining the findings of the two stages (Stage 1: 3 + Stage 2: 25), we totaled 28 studies, presented in Table \ref{tbl:related} and briefly described below.

In order to summarize the related literature presented above and answer our five RQs we extracted data in three stages. During Stage 1, we extracted the data and completed the associated forms, the content of which is summarized in Table \ref{table:extraction}. Then, in Stage 2, we grouped the data by topics. Finally, in Stage 3, we constructed summaries and visual representations of the data to help us analyze and classify the use of EBSE as a research instrument for students.

\begin{table*}[ht]
	\caption{Summary of the extracting forms} \label{table:extraction}
	\vskip 0.3cm
	\centering
	\begin{tabular}{|c||p{9cm}|}
		\hline
		\textbf{Category} & \textbf{Rationale} \\ \hline \hline
		\textbf{Paper metadata} & Used to manage and identify \textbf{WHEN}/\textbf{WHERE} the studies were published -- RQ1\\ \hline
		\textbf{EBSE application} & Used to identify \textbf{HOW} EBSE has been used, for \textbf{WHAT} purpose  and by \textbf{WHOM} -- RQ2 \\ \hline
		\textbf{EBSE benefits} & Used to identify the advantages in using EBSE in an academic context -- RQ3 \\ \hline
		\textbf{EBSE problems} & Used to identify the disadvantages in using EBSE in an academic context -- RQ4 \\ \hline
		\textbf{Other information} & Additional information could be recorded in this field \\ \hline		
	\end{tabular}
\end{table*}

We considered different facets for classifying the studies. The facets were defined according to the RQs and taking the selected studies into account. So, depending on the focus of each scheme, the study was classified as either one or as any combination. In the following, the categories of these facets are presented. 

The categories were established during data extraction, based on data provided by the analyzed studies. First of all, we read the full text of all included studies to identify concepts that reflected the contribution of each study. Secondly, we combined the set of concepts to obtain representative categories. 

\begin{itemize}
	\item \textbf{Purposes (RQ2)}: in this facet, we wanted to learn the purposes for applying EBSE in SE. We have identified five main categories of such purposes:
	\subitem $\rightarrow$ \textbf{Assistance in Teaching and Learning}: this classification is related to the initiatives that use EBSE as an educational tool in SE;
	\subitem $\rightarrow$ \textbf{Benefits and Problems in conducting secondary studies}: this classification focuses on studies that explore students' opinions about their experiences of conducting secondary studies, mainly the identification of benefits and problems encountered;
	\subitem $\rightarrow$ \textbf{Including EBSE into the curricula}: this classification covers studies discussing on how to include EBSE into the curriculum and its challenges;
	\subitem $\rightarrow$ \textbf{Assessing the use of secondary studies}: this classification represents studies investigating applicability, consistency and reliability of secondary studies processes and its application; and
	\subitem $\rightarrow$ \textbf{Application of secondary studies in actual projects}: this classification summarizes how the findings of a secondary study can guide research efforts in actual research projects.\\
	
	\item \textbf{Reported benefits (RQ3)}: the categories for this facet are based on the main benefits related to the use of EBSE in an academic context reported in the included studies. We have identified 10 main categories of benefits, namely: 
	
	\subitem $\rightarrow$ Developing research skills; 
	\subitem $\rightarrow$ Encouraging students to reflect; 
	\subitem $\rightarrow$ Helping find out research opportunities; 
	\subitem $\rightarrow$ Providing a good overview of the literature; 
	\subitem $\rightarrow$ Perceiving the importance of searching for evidence; 
	\subitem $\rightarrow$ Acquiring experience of reading and understanding scientific studies; 
	\subitem $\rightarrow$ Learning from studies and getting knowledge; 
	\subitem $\rightarrow$ Publishing a paper; 
	\subitem $\rightarrow$ Providing baselines to assist new research efforts; and 
	\subitem $\rightarrow$ Collaborating with other researchers (teamwork).\\
	
	\item \textbf{Reported Problems (RQ4)}: the categories for this facet are based on the main problems/difficulties highlighted by the authors of the included studies related to use of EBSE in an academic context in SE. We have identified 13 main categories of problems, namely:
	\subitem $\rightarrow$ Lacking of domain knowledge;
	\subitem $\rightarrow$ Lacking of SLR/SMS process knowledge;
	\subitem $\rightarrow$ Developing the protocol is longer than expected;
	\subitem $\rightarrow$ Defining research questions; 
	\subitem $\rightarrow$ Finding synonyms and keywords; 
	\subitem $\rightarrow$ Searching the literature; 
	\subitem $\rightarrow$ Defining and applying the inclusion/exclusion criteria; 
	\subitem $\rightarrow$ Appraising the evidence for its validity; 
	\subitem $\rightarrow$ Extracting data; 
	\subitem $\rightarrow$ Classifying the literature; 
	\subitem $\rightarrow$ Extending the period of time to perform the study; 
	\subitem $\rightarrow$ Writing the report (SLR's results); and 
	\subitem $\rightarrow$ Limiting space while publishing.\\
	
\end{itemize}	     

\subsection{Results}

Here we present the results of our mapping as we used the facets of the classification schema aforementioned to answer the RQs. 

\subsubsection{Frequency of Publication (RQ1)}

In order to offer a general view of the efforts when using EBSE in an academic context in SE, a distribution of the 28 selected studies over the years is shown in Table \ref{table:Year}. As it suggests, the research of EBSE in SE is recent, secondary studies were recognized as an EBSE method since Kitchenham et al. \cite{Kitchenham04a} have published their pioneer work. The interest of its use as an educational tool has been moderately increasing within 2009--2013, without a significant increase during other years.

\begin{table}[!ht]
	\caption{Frequency of studies per year} \label{table:Year}
	\centering
	\begin{tabular}{|c||l|}
		\hline
		\textbf{Year} & \textbf{Total of studies -- Studies ID} \\ \hline \hline
		\textbf{2005} & 1 -- S1 \cite{Jorgensen05} \\ \hline
		\textbf{2006} & 1 -- S2 \cite{Rainer06} \\ \hline
		\textbf{2007} & 1 -- S3 \cite{Wohlin07} \\ \hline
		\textbf{2008} & 2 -- S4 \cite{Baldassarre08}; S5 \cite{Rainer08} \\ \hline
		\textbf{2009} & 4 -- S6 \cite{Brereton09}; S7 \cite{Janzen09}; S8 \cite{Oates09}; S9 \cite{Babar09} \\ \hline
		\textbf{2010} & 3 -- S10 \cite{Riaz10}; S11 \cite{MacDonell10}; S12 \cite{Kitchenham10b} \\ \hline		
		\textbf{2011} & 3 -- S13 \cite{Kitchenham11a}; S14 \cite{Brereton11}; S15 \cite{Kitchenham11} \\ \hline		
		\textbf{2013} & 4 -- S16 \cite{Carver13}; S17 \cite{Castelluccia13}; S18 \cite{Catal13}; S19 \cite{Wohlin13} \\ \hline		
		\textbf{2014} & 2 -- S20 \cite{Dekhane14}; S21 \cite{Lavallee14} \\ \hline		
		\textbf{2015} & 3 -- S22 \cite{Clear15}; S23 \cite{Pejcinovic15}; S24 \cite{Souza15a} \\ \hline
		\textbf{2017} & 2 -- \textcolor{black}{S25} \cite{Kaijanaho17}; S26 \cite{Kuhrmann17} \\ \hline
		\textbf{2018} & 2 -- S27 \cite{Molleri18}; S28 \cite{Ribeiro18} \\ \hline
	\end{tabular}
\end{table}

Looking at the publication vehicle (see Table \ref{table:Publication}), conferences seem to be the main communication channel representing 53.5\% (15 studies) of publications. Journals represent 32.1\% (9 studies) and finally symposiums with 14.2\% (4 studies).

\begin{table*}[ht]
	\caption{Publication vehicle} \label{table:Publication}
	\vskip 0.3cm
	\centering
	\begin{tabular}{|c||p{9cm}|}
		\hline
		\textbf{Publication Venue} & \textbf{Total of studies -- Studies ID} \\ \hline \hline
		
		\textbf{Conference} & 15 (53.5\%) -- S2 \cite{Rainer06}; S3 \cite{Wohlin07}; S4 \cite{Baldassarre08}; S5 \cite{Rainer08}; S6 \cite{Brereton09}; S8 \cite{Oates09};  S10 \cite{Riaz10}; S12 \cite{Kitchenham10b}; S13 \cite{Kitchenham11a};  S22 \cite{Clear15}; S23 \cite{Pejcinovic15}; S24 \cite{Souza15a}; S25 \cite{Kaijanaho17}; S26 \cite{Kuhrmann17}; and S27 \cite{Molleri18}  \\ \hline
		
		\textbf{Journal} & 9 (32.1\%) -- S7 \cite{Janzen09}; S11 \cite{MacDonell10}; S14 \cite{Brereton11}; S15 \cite{Kitchenham11}; S17 \cite{Castelluccia13}; S18 \cite{Catal13}; S19 \cite{Wohlin13}; S21 \cite{Lavallee14}; and S28 \cite{Ribeiro18}   \\ \hline
		
		\textbf{Symposium} & 4 (14.2\%) -- S1 \cite{Jorgensen05}; S9 \cite{Babar09}; S16 \cite{Carver13}; and S20 \cite{Dekhane14} \\ \hline
		
	\end{tabular}
\end{table*}

\subsubsection{Purposes of EBSE in an academic context (RQ2)}

We can notice that studies involving undergraduate students (12 (42.8\%) -- S1 \cite{Jorgensen05}; S2 \cite{Rainer06}; S3 \cite{Wohlin07}; S5 \cite{Rainer08}; S6 \cite{Brereton09}; S12 \cite{Kitchenham10b}; S13 \cite{Kitchenham11a}; S14 \cite{Brereton11}; S20 \cite{Dekhane14}; S21 \cite{Lavallee14}; S22 \cite{Clear15}; S25 \cite{Kaijanaho17}) have the largest representativeness. The other two categories related to students were: $\bullet$ Master students (11 (39.2\%) -- S4 \cite{Baldassarre08}; S7 \cite{Janzen09}; S8 \cite{Oates09}; S16 \cite{Carver13}; S17 \cite{Castelluccia13}; S18 \cite{Catal13}; S23 \cite{Pejcinovic15}; S25 \cite{Kaijanaho17}; S26 \cite{Kuhrmann17}; S27 \cite{Molleri18}; S28 \cite{Ribeiro18}); and $\bullet$ Ph.D. students (8 (28.5\%) -- S9 \cite{Babar09}; S10 \cite{Riaz10}; S12 \cite{Kitchenham10b}; S15 \cite{Kitchenham11}; S16 \cite{Carver13}; S24 \cite{Souza15a}; S27 \cite{Molleri18}; S28 \cite{Ribeiro18}). Only 4 studies (14.2\%) assessed the use of EBSE by SLR experts (S10 \cite{Riaz10}; S11 \cite{MacDonell10}; S15 \cite{Kitchenham11}; S19 \cite{Wohlin13}). Some categories that were identified are strongly related, for instance, undergraduate and Masters (e.g. S16 \cite{Carver13}); Masters and Ph.Ds. (S25 \cite{Kaijanaho17}); or SLR experts and Ph.D. students (e.g., S15 \cite{Kitchenham11}), thus, these studies were considered in more than one category.

In summary, while there have been several prior investigations on undergraduate (12 studies -- 42.8\%) and graduate levels -- M.Sc./Ph.D. (19 studies -- 67.7\%), few focus on SE experts (4 studies -- 14.2\%). Table \ref{tbl:related} shows different aspects of the adoption of secondary studies in SE. In the sequence, these aspects, which were grouped in 5 classifications, are discussed.

\begin{table}[!htbp]
	\centering
	\caption{Summary of Prior Studies on the use/contribution of secondary studies in SE}
	\label{tbl:related}
	\begin{tabular}{|c||p{6cm}|p{6.5cm}|p{0.8cm}|} \hline 
		\textbf{{\small ID}} & \textbf{{\small Purpose}} & \textbf{{\small Main topic}} & \textbf{{\small Ref.}} \\ [0.1pt] \hline \hline
		
		{\small \textbf{S1}} & {\small Undergraduate students} & {\small Experiences about teaching EBSE} & {\small \cite{Jorgensen05}} \\ [0.1pt] \hline 	
		
		{\small \textbf{S2}} & {\small Undergraduate students} & {\small Teaching Empirical Methods to Undergraduate Students} & {\small \cite{Rainer06}} \\ [0.1pt] \hline
		
		{\small \textbf{S3}} & {\small Undergraduate students} & {\small Introducing  EBSE into a course/curricula} & {\small \cite{Wohlin07}} \\ [0.1pt] \hline 																	
		
		{\small \textbf{S4}} & {\small Masters students} & {\small Experiences about teaching EBSE} & {\small \cite{Baldassarre08}} \\ [0.1pt] \hline 		
		
		{\small \textbf{S5}} & {\small Undergraduate students} & {\small The use of EBSE by undergraduate students} & {\small \cite{Rainer08}} \\ [0.1pt] \hline
		
		{\small \textbf{S6}} & {\small Undergraduate students} & {\small The applicability of the SLR process}  & {\small \cite{Brereton09}} \\ [0.1pt] \hline 
		
		{\small \textbf{S7}} & {\small Masters students} & {\small EBSE web-database} & {\small \cite{Janzen09}} \\ [0.1pt] \hline
		
		{\small \textbf{S8}} & {\small Masters students} & {\small Research skills} & {\small \cite{Oates09}} \\ [0.1pt] \hline
		
		{\small \textbf{S9}} & {\small Ph.D. students} & {\small Value of SLRs for Ph.D. students} & {\small \cite{Babar09}} \\ [0.1pt] \hline 	
		
		{\small \textbf{S10}} & {\small SLR experts and Ph.D. students} & {\small Experiences conducting SLRs} & {\small \cite{Riaz10}} \\ [0.1pt] \hline 		
		
		{\small \textbf{S11}} & {\small SLR experts} & {\small Stability of SLR outcomes} & {\small \cite{MacDonell10}} \\ [0.1pt] \hline 		
		
		{\small \textbf{S12}} & {\small Undergraduate and Ph.D. students} & {\small Educational and scientific value of SM} & {\small \cite{Kitchenham10b}} \\ [0.1pt] \hline 	
		
		{\small \textbf{S13}} & {\small Undergraduate students} & {\small Repeatability of SLRs} & {\small \cite{Kitchenham11a}} \\ [0.1pt] \hline 					
		
		{\small \textbf{S14}} & {\small Undergraduate students} & {\small Effectiveness of SLRs} & {\small \cite{Brereton11}} \\ [0.1pt] \hline 		
		
		{\small \textbf{S15}} & {\small SLR experts and Ph.D. students} & {\small Value of SMs} & {\small \cite{Kitchenham11}} \\ [0.1pt] \hline 				
		
		{\small \textbf{S16}} & {\small Masters and Ph.D. students} & {\small Mentors for SLR conduction} & {\small \cite{Carver13}} \\ [0.1pt] \hline 																	
		
		{\small \textbf{S17}} & {\small Masters students} & {\small Experiences about teaching EBSE} & {\small \cite{Castelluccia13}} \\ [0.1pt] \hline 															
		
		{\small \textbf{S18}} & {\small Masters students} & {\small Experiences about teaching EBSE} & {\small \cite{Catal13}} \\ [0.1pt] \hline 	
		
		{\small \textbf{S19}} & {\small SLR experts} & {\small Reliability of SM} & {\small \cite{Wohlin13}} \\ [0.1pt] \hline 		
		
		{\small \textbf{S20}} & {\small Undergraduate students} & {\small Experiences about teaching EBSE} & {\small \cite{Dekhane14}} \\ [0.1pt] \hline 						
		
		{\small \textbf{S21}} & {\small Undergraduate students} & {\small Variation of the traditional SLR process -- Interactive SR} & {\small \cite{Lavallee14}} \\ [0.1pt] \hline 		
		
		{\small \textbf{S22}} & {\small Undergraduate students} & {\small Experience about teaching EBSE} & {\small \cite{Clear15}} \\ [0.1pt] \hline 		
		
		{\small \textbf{S23}} & {\small Masters students} & {\small Experience about teaching EBSE} & {\small \cite{Pejcinovic15}} \\ [0.1pt] \hline 		
		
		{\small \textbf{S24}} & {\small Ph.D. students} & {\small Use of SM findings} & {\small \cite{Souza15a}} \\ [0.1pt] \hline 		
		
		{\small \textbf{S25}} & {\small Undergraduate and Masters students} & {\small Skill for reading research literature} & {\small \cite{Kaijanaho17}} \\ [0.1pt] \hline 													
		{\small \textbf{S26}} & {\small Masters students} & {\small Experiences about teaching EBSE} & {\small \cite{Kuhrmann17}} \\ [0.1pt] \hline 																		
		
		{\small \textbf{S27}} & {\small Masters and Ph.D. students} & {\small Critically appraise scientific literature} & {\small \cite{Molleri18}} \\ [0.1pt] \hline 																			
		
		{\small \textbf{S28}} & {\small Masters and Ph.D. students} & {\small Reliability of SLRs} & {\small \cite{Ribeiro18}} \\ [0.1pt] \hline 																			
	\end{tabular}
\end{table}

\textbf{Assistance in Teaching and Learning}: 11 studies (39.2\%) used secondary studies as a means for teaching and learning undergraduate/graduate (M.Sc./Ph.D.) courses in SE; representing that secondary studies are effective educational tools.

\begin{itemize}
	\item \textbf{S1} -- J{\o}rgensen et al. \cite{Jorgensen05}, based on their previous experience and lessons learned in teaching EBSE to undergraduate students, wanted to inspire and support other universities to develop their own EBSE courses;
	\item \textbf{S4} -- In \cite{Baldassarre08}, Baldassarre et al. described how SLRs were integrated as part of an EBSE course through a hands-on approach. According to the authors their approach was a useful means for making Masters students understand the importance of searching for evidence using SLR;
	\item \textbf{S7} -- Janzen and Ryoo \cite{Janzen09} idealized and developed a community-driven Web database containing summaries of EBSE studies. This database was integrated into activities in a Master EBSE course;
	\item \textbf{S8} -- Oates and Capper \cite{Oates09} addressed the problem of teaching research skills to Masters students. They suggested to introduce SLRs and EBSE to improve, for example, students' literature-handling skills and well-defined research question formulation. They concluded that Masters students could be benefited by SLRs;
	\item \textbf{S16} -- Kuhrmann \cite{Kuhrmann12, Kuhrmann17} described his experience in teaching EBSE using expert teams in a Master-level course enrolled with 70 students. In this course a seminar-like learning model where students form expert teams was used. These teams obtain extra expertise that they offer as service to other teams (cross-team collaboration);
	\item \textbf{S18} -- Castelluccia et al. \cite{Castelluccia13}  shared their experiences about teaching EBSE to Masters students in Computer Science. The students conducted a collaborative SMS on challenges in the adoption of open source software in a business context. The results were positive in terms of students' participation, teamwork attitude, professional interest in open source software, and exam passing;
	\item \textbf{S19} -- Catal \cite{Catal13} also described experiences on teaching EBSE to Master students. Each of the students delivered an SMS on software architecture. The results showed that the teaching approach was quite useful for Masters students; 
	\item \textbf{S22} -- Lavall\'ee et al. \cite{Lavallee14} advocated using SLR as an educational tool for undergraduate students. They created a variation of the traditional SLR process called Interactive SR (iSR). In this adapted process the authors proposed to refine the protocol in iterations. Using iSR to teach SLR for undergraduate students could be beneficial because in each iteration, there is a progressive gain in expertise. Moreover, there is an amount of SLR's skills acquired in each iteration. Another advantage of an iterative approach is the refinement opportunity; 
	\item \textbf{S23} -- Clear \cite{Clear15} adopted the SLR process to an undergraduate teaching setting aiming to produce an SLR on challenges and recommendations for the designing and conducting global software engineering courses. The author concluded that SLRs definitely have value as a method for research and teaching;
	\item \textbf{S24} -- In \cite{Pejcinovic15}, Pejcinovic mentioned that the practice in SLR could produce skills for undergraduate students, such as, improvement of critical thinking, better writing, increasing motivation, among others. He reinforced that the advantages of SLR in undergraduate education level are clear; and
	\item \textbf{S26} -- Kaijanaho \cite{Kaijanaho17} has attempted to teach the skill to read research literature critically to Masters students in a master's degree level course in programming languages over 15 years. He discussed his experience on the use of evidence-based practice and a flipped classroom for this purpose. Kaijanaho presented no firm conclusions,  however he defends that university students, even at the bachelor's and master's degrees phases, should be given EBSE principles for their academic training.
\end{itemize}

\textbf{Benefits and Problems in conducting secondary studies}: Looking at the studies that presented lessons learned, including benefits and problems in conducting secondary studies, 6 studies were identified (21.4\%).

\begin{itemize}
	\item \textbf{S5} -- Rainer et al. \cite{Rainer08} evaluated the use of EBSE by undergraduate students. The students that were interviewed found EBSE very challenging, in particular the SLRs;		
	\item \textbf{S9} -- Babar and Zhang \cite{Babar09} conducted a survey to explore opinions of research practitioners about their experiences in conducting SLRs. One of the analyzed perspectives was the value of SLRs for first-year Ph.D. students. Participants pointed out that some advantages of SLRs for Ph.D. students include: (i) systematic way of building a body of knowledge about a particular topic or research question (what has been done); (ii) clear statement and structure of state-of-the-art; (iii) learn from studies and getting knowledge; and (iv) find out gaps -- research opportunities (ideas for further research);
	\item \textbf{S10} -- Riaz et al. \cite{Riaz10} reported experiences of expert and non-expert researchers in conducting SLRs. The experiences and problems faced by expert researchers and non-experts (Ph.D. students) while conducting SLRs, and the approaches adopted to tackle the issues encountered vary. In summary, it is observed that Ph.D. students can be faced with big challenges, mainly due to their limited experience with SLR and, sometimes, with the research topic. The main problem faced by Ph.D. students is related to defining quality criteria;
	\item \textbf{S12} -- The educational and scientific value of undergraduate and graduate (M.Sc./Ph.D.) students undertaking SMS was assessed by Kitchenham et al. \cite{Kitchenham10b}. Students had a valuable experience undertaking an SMS. As positive aspects they mentioned that SMS provides reusable research skills and a good overview of a research topic. However, undertaking the study in the required timescales, searching and classifying the literature, were the reported problems;
	\item \textbf{S14} -- A case study was conducted by Brereton \cite{Brereton11} to explore the effectiveness of second-year undergraduate computing students in carrying out an SLR. The results suggest that the students found the conducting phase, including the selection activity, more problematic than the planning phase. Pearl concluded that undergraduates can perform SLRs (specially if undertaken by groups), but the task is clearly quite challenging and time-consuming; and	
	\item \textbf{S15} -- In order to assess the value of SMSs Kitchenham et al. \cite{Kitchenham11} use a multi-case, participant-observer case study using five studies that were preceded by SMS. The research question addressed by this case study was: ``How do mapping studies contribute to further research?'' As a result, Kitchenham et al. have identified some benefits that can accrue from basing research on a preceding SMS, e.g., SMS can save time and effort for researchers and also provides baselines to assist new research efforts.
\end{itemize}

\textbf{Including EBSE into the curricula}: Even with a smaller percentage of studies, 2 studies (7.1\%) discussed how to include EBSE into the SE curricula.

\begin{itemize}
	\item \textbf{S3} -- Wohlin \cite{Wohlin07} presented several aspects and challenges when introducing EBSE into the curricula. A positive aspect of this introduction into the curricula is the possibility to run empirical studies in student settings. A challenge is to balance educational and research objectives; and
	\textbf{S21} -- \cite{Dekhane14} described how to incorporate research, professional practice and methodologies into two undergraduate SE courses. The courses enabled introduction to undergraduate students on basic research skills.	
\end{itemize}

\textbf{Assessing the use of secondary studies}: 8 studies (28.5\%) have investigated EBSE more generally, emphasizing secondary studies' processes and the reliability of the results produced by their use.
\begin{itemize}
	\item \textbf{S2} -- Rainer et al. \cite{Rainer06} reported the use of EBSE by 15 final-year undergraduate students. The investigation produced inconsistent results. On one hand, their quantitative data suggested that students are making good use of some of the EBSE guidelines. On the other hand, their qualitative evidence suggested that the students are not using the guidelines properly;
	\item \textbf{S6} -- Brereton et al. \cite{Brereton09} investigated the applicability of the SLR process by undergraduate students over a relatively short period of time. In common, they found that, with certain modifications to the process, it was possible to perform an SLR within a limited time period and to generate valid results;
	\item \textbf{S11} -- MacDonell et al. \cite{MacDonell10} investigated the consistency of SLR process and the stability of outcomes. The authors found out that two independent SLRs were very similar when produced by domain experts with experience on the SLR process;
	\item \textbf{S13} -- Kitchenham et al. \cite{Kitchenham11a} undertook a study to investigate the repeatability of SLRs (i.e., identification of the same studies when SLRs are performed independently by two novice researchers -- undergraduate students). They concluded that in the case of novice researchers (undergraduate students) they will not necessarily select the same studies, in other words, they will not guarantee repeatability with respect to studies;
	\item \textbf{S17} -- Carver et al. \cite{Carver13} suggested how to help mentors guide Masters and Ph.D. students through the SLR process. Their findings highlight the importance of mentoring by an experienced researcher throughout the process;
	\item \textbf{S20} -- Wohlin et al. \cite{Wohlin13} also evaluated the reliability of SMS. They investigated two SMSs on software product line testing. They concluded that despite both studies are addressing the same topic, there are differences in terms of the number of included papers and their classification (categorization); 
	\item \textbf{S27} -- Molleri \cite{Molleri18} investigated the application of checklists for supporting M.Sc./Ph.D. students to critically appraise scientific literature in a postgraduate EBSE course. 76 students (in pairs) participated in the experiment. They used two checklists to evaluate two papers (reporting a case study and an experiment) each. Students perceived checklist items as difficult to assess.The conclusion was that the clearer the reporting, the easier it was for students to judge the quality of studies; and
	\item \textbf{S28} -- Ribeiro et al. \cite{Ribeiro18} mention that if similar SLR protocols are executed by similar teams of Masters and Ph.D. students, they should lead to equivalent answers for the same research question. The outcomes were different and six challenges contributing to the divergences, e.g.,  researchers' inexperience in the topic and researchers' inexperience in the method, among others. According to Ribeiro et al. it is not possible to rely on results of SLRs performed by Masters and Ph.D. students. 		
\end{itemize}

\textbf{Application of secondary studies in real projects}: Only 1 study (3.5\%) discussed how the findings of a secondary study guided research efforts.

\begin{itemize}
	\item \textbf{S25} -- In summary, Souza and others \cite{Souza15a} argue that SMS had a great importance in conducting their Ph.D. projects. For example, from the findings of a mapping on Knowledge Management (KM) in Software Testing they noticed that the application of KM strategies in the field of software testing was a very promising research area. Moreover, the mapping also showed them that ontology-based KM solutions were even rarer in the software testing domain. This finding attracted their attention, since ontology is recognized as an important instrument for supporting KM in general. After the SMS, they performed an SLR to investigate in details ontology-based KM in software testing. Finally, they undertook an on-line survey on the most important aspects of KM when applied to software testing. They used insights from the mapping to formulate some of the survey questions. Among the benefits of using SMS, they can highlight that: (i) it establishes a solid baseline that serves as an important guide for research efforts, e.g. it shows gaps in a research topic; (ii) the identification of gaps is essential for definition of research questions to be investigated, and to define research strategies to follow; and (iii) it identifies several studies, which can be used as a baseline for comparison.
\end{itemize}

\subsubsection{Reported Benefits (RQ3) and Problems Using EBSE in an academic context (RQ4)}

As we can see from this study (Table \ref{table:Benefits}), there are 10 benefits of using EBSE in an academic context in SE. These benefits were named as B1--B10 in Table \ref{table:Benefits}. 
\textcolor{black}{Several studies (S5, S9, S10, S12, S14, S15) reported that secondary studies have been used to provide future SE researchers/practitioners with the knowledge and skills to enable them to make decisions based on evidence. Some of the scientific methods to become skillful are developed while conducting secondary studies, for example, identifying, selecting, interpreting, and evaluating evidence. These skills enable both future researchers and practitioners to make better decisions.} 

\begin{table*}[!ht]
	\caption{Identified benefits using EBSE in an academic context} \label{table:Benefits}
	\vskip 0.3cm
	\centering
	\begin{tabular}{|p{6cm}||p{7cm}|} 
		\hline
		\textbf{Benefit} & \textbf{Total of studies -- Studies ID} \\ \hline \hline		
		\textbf{(B1)} Developing research skills & \small \textbf{5} -- S1 \cite{Jorgensen05}; S2 \cite{Rainer06}; S8 \cite{Oates09}; S20 \cite{Dekhane14}; S22 \cite{Clear15} \\ \hline
		\textbf{(B2)} Encouraging students to reflect & \small \textbf{5} -- S2 \cite{Rainer06}; S8 \cite{Oates09}; S14 \cite{Brereton11}; S24 \cite{Souza15a}; S25 \cite{Kaijanaho17} \\ \hline
		\textbf{(B3)} Helping find out research opportunities & \small \textbf{3} -- S9 \cite{Babar09}; S15 \cite{Kitchenham11}; S24 \cite{Souza15a} \\ \hline
		\textbf{(B4)} Providing an overview of the literature & \small \textbf{5} -- S12 \cite{Kitchenham10b}; S15 \cite{Kitchenham11}; S20 \cite{Dekhane14}; S22 \cite{Clear15}; S24 \cite{Souza15a} \\ \hline
		\textbf{(B5)} Perceiving the importance of searching for evidence & \small \textbf{6} -- S4 \cite{Baldassarre08}; S5 \cite{Rainer08}; S7 \cite{Janzen09}; S8 \cite{Oates09}; S9 \cite{Babar09}; S22 \cite{Clear15} \\ \hline
		\textbf{(B6)} Acquiring experience of reading/understanding scientific studies & \small \textbf{5} -- S1 \cite{Jorgensen05}; S2 \cite{Rainer06}; S8 \cite{Oates09}; S15 \cite{Kitchenham11}; S25 \cite{Kaijanaho17} \\ \hline
		\textbf{(B7)} Learning from studies and getting knowledge & \small \textbf{4} -- S9 \cite{Babar09}; S15 \cite{Kitchenham11}; S20 \cite{Dekhane14}; S25 \cite{Kaijanaho17} \\ \hline
		\textbf{(B8)} Publishing a paper & \small \textbf{1} -- S9 \cite{Babar09} \\ \hline
		\textbf{(B9)} Providing baselines to assist new research efforts & \small \textbf{3} -- S15 \cite{Kitchenham11}; S15 \cite{Kitchenham11}; S24 \cite{Souza15a} \\ \hline
		\textbf{(B10)} Collaborating with other researchers (teamwork) & \small \textbf{2} -- S17 \cite{Castelluccia13}; S26 \cite{Kuhrmann17} \\ \hline		
	\end{tabular}
\end{table*}

Despite the use of secondary studies bringing many benefits, some studies confirmed that adopting secondary studies in SE is challenging and a total of 13 problems (difficulties), named as P1--P13 in Table \ref{table:Problems}, were identified. 
\textcolor{black}{We observed that while some of the challenges for those conducting secondary studies are common to both expert researchers and students, e.g., searching the literature (S5, S9, S11, S12, S13, S16, S18, S19, S23); and appraising the evidence for its validity and extracting data (S5, S16, S18, S27), there are others, e.g., developing the protocol (S6, S9, S10, S21, S23); finding synonyms and keywords (S10, S13, S16, S21, S18); and defining and applying inclusion/exclusion criteria (S6, S10, S12, S13, S14, S16, S19, S21, S27), that are only faced by students due to their limited experience with secondary studies in particular and with research in general.}

\begin{table*}[!ht]
	\caption{Identified problems using EBSE in an academic context} \label{table:Problems}
	\vskip 0.3cm
	\centering
	\begin{tabular}{|p{6cm}||p{7cm}|} 
		\hline
		\textbf{Problems (P)} & \textbf{Total of studies -- Studies ID} \\ \hline \hline
		\textbf{(P1)} Lacking of domain knowledge & \small \textbf{2} -- S9 \cite{Babar09}; S28 \cite{Ribeiro18} \\ \hline
		\textbf{(P2)} Lacking of SLR/SM process knowledge & \small \textbf{1} -- S28 \cite{Ribeiro18} \\ \hline
		\textbf{(P3)} Developing the protocol is longer than expected &  \small \textbf{5} -- S6 \cite{Brereton09}; S9 \cite{Babar09}; S10 \cite{Riaz10}; S21 \cite{Lavallee14}; S23 \cite{Pejcinovic15} \\ \hline
		\textbf{(P4)} Defining research questions & \small \textbf{4} -- S9 \cite{Babar09}; S10 \cite{Riaz10}; S16 \cite{Carver13}; S21 \cite{Lavallee14} \\ \hline 
		\textbf{(P5)} Finding synonyms and keywords & \small \textbf{5} -- S10 \cite{Riaz10}; S13 \cite{Kitchenham11a}; S16 \cite{Carver13}; S21 \cite{Lavallee14}; S28 \cite{Ribeiro18} \\ \hline
		\textbf{(P6)} Searching the literature & \small \textbf{9} -- S5 \cite{Rainer08}; S9 \cite{Babar09}; S11 \cite{MacDonell10}; S12 \cite{Kitchenham10b}; S13 \cite{Kitchenham11a}; S16 \cite{Carver13}; S18 \cite{Catal13}; S19 \cite{Wohlin13}; S23 \cite{Pejcinovic15} \\ \hline
		\textbf{(P7)} Defining and applying the inclusion/exclusion criteria & \small \textbf{9} -- S6 \cite{Brereton09}; S10 \cite{Riaz10}; S12 \cite{Kitchenham10b}; S13 \cite{Kitchenham11a}; S14 \cite{Brereton11}; S16 \cite{Carver13}; S19 \cite{Wohlin13}; S21 \cite{Lavallee14}; S27 \cite{Molleri18} \\ \hline
		\textbf{(P8)} Appraising the evidence for its validity & \small \textbf{4} -- S5 \cite{Rainer08}; S16 \cite{Carver13}; S18 \cite{Catal13}; S27 \cite{Molleri18} \\ \hline
		\textbf{(P9)} Extracting data & \small \textbf{5} -- S14 \cite{Brereton11}; S15 \cite{Kitchenham11}; S16 \cite{Carver13}; S21 \cite{Lavallee14}; S23 \cite{Pejcinovic15}  \\ \hline
		\textbf{(P10)} Classifying the literature & \small \textbf{5} -- S12 \cite{Kitchenham10b}; S19 \cite{Wohlin13}; S21 \cite{Lavallee14}; S23 \cite{Pejcinovic15}; S27 \cite{Molleri18}  \\ \hline
		\textbf{(P11)} Extended period of time to perform the study & \small \textbf{4} -- S5 \cite{Rainer08}; S10 \cite{Riaz10}; S12 \cite{Kitchenham10b}; S23 \cite{Pejcinovic15} \\ \hline
		\textbf{(P12)} Writing the report (SLR's results) & \small \textbf{2} -- S23 \cite{Pejcinovic15}; S27 \cite{Molleri18} \\ \hline
		\textbf{(P13)} Limited space while publishing & \small \textbf{2} -- S10 \cite{Riaz10}; S14 \cite{Brereton11} \\ \hline
	\end{tabular}
\end{table*}

The main contribution of this SMS is to provide evidence on how secondary studies have been used in SE academic contexts. It is clear that secondary studies are not only being conducted by experienced researchers but also by students, including Masters and Ph.D. students. 

A group of researchers have investigated the reliability of secondary studies as a research instrument (S2, S6, S11, S13, S17, S20, S27, S28). They concluded that the level of experience in secondary studies and research influences the repeatability of both process and outcomes. The results suggested that undergraduate/graduate (M.Sc./Ph.D.) students can perform secondary studies, especially if undertaken by groups, and under the supervision of an expert.

It is clear from the above mentioned studies that, despite the difficulties, there are advantages in adopting secondary studies. However, it is not clear how a secondary study may contribute to M.Sc. or Ph.D. research project in the SE area. Our results revealed that there is only one research study (S25) \cite{Souza15a} clarifying this topic. Moreover, we believe it would be very helpful to know the relationship between secondary studies and activities at M.Sc./Ph.D. level. Therefore, due to the lack of studies and the importance of this issue, we decided to conduct a survey to verify, in a more specific way, the use of secondary studies on behalf of graduate students (Master/PhD) and their perceptions on it. 

\section{Survey on the application of secondary studies in research projects} \label{survey}

In the second part of this study, we conducted a survey on the application of secondary studies in research projects. We followed the six phases proposed by Kitchenham and Pfleeger \cite{Kitchenham08b} to conduct surveys:

\textbf{-- Phase 1}: Setting the objectives --  Our objective can be described by the following RQ: ``What is the practical application of secondary studies in graduate (M.Sc./Ph.D.) research projects?''

\textbf{-- Phase 2:} Designing the survey -- As occurs with most of the surveys in SE \cite{Kitchenham08}, our survey is also a cross-sectional study. Participants were asked about their past experiences on using secondary study results to support decisions on their research project.

\textbf{-- Phase 3:} Developing the survey instrument (i.e., the questionnaire) -- The survey comprises a questionnaire with 4 sections: 1) Profile (3 questions); 2) Practical view on the use of secondary studies (3 questions); 3) Students/researchers' perception on the use of secondary studies (10 agree/disagree questions); and 4) Final comments (1 question). The questions of Section 2 refer to the same RQs answered in the SMS (see Section \ref{relatedWorks}), allowing a confrontation of the theoretical evidences with the practical perspective on the use of secondary studies in an academic context. For example, the answer options for the question: ``What are the main benefits of conducting a secondary study during a research project?'' are the benefits summarized in Table \ref{table:Benefits}. An on-line version of the survey is available on: \textit{https://goo.gl/Bj7FWm}.

The possible choices to the agree/disagree questions vary from ``Strongly agree'' to ``Strongly disagree'', in a scale based on the Likert Scale method, which is a metric used in questionnaires such as attitude surveys. The final median score represents overall level of accomplishment or attitude toward the subject matter. Moreover, the sequence includes a final comments section where respondents can provide any comments. 

\textbf{-- Phase 4:} Evaluating the survey instrument -- We conducted a pre-test, i.e. we applied the survey to a smaller sample (3 SE researchers), intending to identify any problems with the questionnaire. The questionnaire was also evaluated by one experienced researcher in secondary studies. One of the modifications was to change the ``Strongly agree'' (option from the Likert Scale) to the right side of the screen and the ``Strongly disagree'' to the left side.

\begin{figure*}[!ht]
	\centering
	\includegraphics [width=1\textwidth]{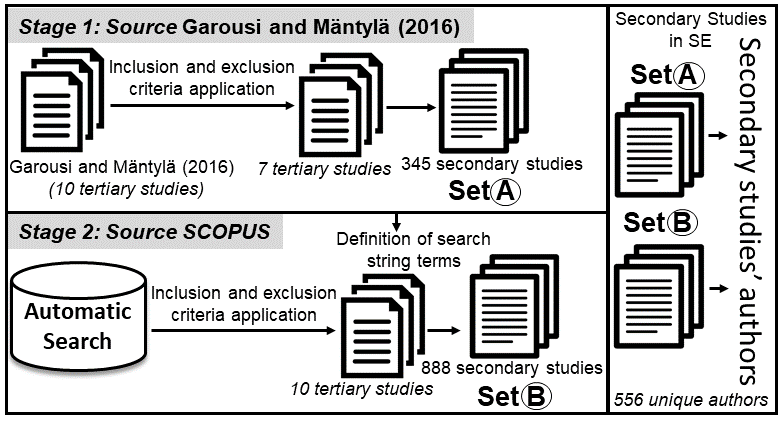}
	\caption{Searching for SE secondary studies' authors}
	\label{Tertiary}
\end{figure*}

\textbf{-- Phase 5:} Obtaining valid data -- Definition of population: In order to identify our population (both experts and novice researchers -- M.Sc. and Ph.D. students -- performing secondary studies) we followed a two-stage systematic search process, as shown in Figure \ref{Tertiary}.

During Stage 1, we reused the list of the 10 tertiary studies identified by Garousi and M{\"a}ntyl{\"a} \cite{Garousi16}. A tertiary study summarizes data from secondary studies, therefore identifying the secondary studies in SE area it was possible to identify researchers who are conducting those studies. Out of the 10 tertiary studies (see Table \ref{tbl:population}) we included 8, totaling 345 secondary studies.

With an objective to expand our set of studies (SE authors of secondary studies), i.e., increase both the number of authors and also the range of publication years, during the second stage, we created a search string (see Table \ref{table:string1}, which was applied in \textit{Scopus} in three metadata fields: title, abstract and keywords. The search string terms definition was based on the reading of titles and abstracts of tertiary studies included in Garousi and M{\"a}ntyl{\"a}' study \cite{Garousi16}.

\begin{table*}[!ht]
	\caption{Keywords of the Search String -- Searching for SE secondary studies' authors} \label{table:string1}
	\vskip 0.3cm
	\centering
	\begin{tabular}{c||p{10cm}}
		\hline
		\textit{\textbf{Search String}} & 
		(``tertiary study'' 						\textbf{OR} 
		``tertiary review'' 						\textbf{OR} 
		``tertiary systematic review''       		\textbf{OR} 
		``systematic review of systematic review'') 
		\textbf{AND}
		``software engineering''
		\\
		\hline
	\end{tabular}
\end{table*}

The selection criteria are organized in two Inclusion Criterion (IC) and seven Exclusion Criteria (EC). The following inclusion criteria were applied:

\begin{itemize}
	\item[]\textbf{(IC1)} The study is a tertiary study and it must be within the SE context \\
	\textbf{(IC2)} The list of secondary studies included in the tertiary study is available.
	
\end{itemize} 

\begin{itemize}
	\item[] \textbf{(EC1)}The study does not have an abstract;\\ 
	\textbf{(EC2)} The study is just published as an abstract;\\ 
	\textbf{(EC3)} The study is not written in English;\\
	\textbf{(EC4)} The study is an older version of other study already considered;\\ 
	\textbf{(EC5)} The study is not a tertiary study, such as editorials, summaries of keynotes, workshops, and tutorials;\\ and
	\textbf{(EC6)} The study is not in the scope of SE, although it is a tertiary study; \\ and
	\textbf{(EC7)} The list of secondary studies included is unavailable, although it is a tertiary study.
	
\end{itemize} 

\begin{table*}[!ht]
	\centering
	\caption{Secondary studies identified by Garousi and M{\"a}ntyl{\"a (2016)}}
	\label{tbl:population}
	\begin{tabular}{|p{5cm}|c|c|c|c|}
		\hline
		\textbf{{\small Study}} & \textbf{{\small Total}} & \textbf{{\small Year}} & \textbf{{\small Ref.}} & \textbf{{\small Status}} \\ [0.1pt] \hline \hline 
		
		{\small SLRs in SE -- An SLR} & {\small 20 secondary studies} & {\small 2009} & {\small \cite{Kitchenham09}} & {\small Included} \\ [0.1pt] \hline 
		{\small SLRs in SE -- A tertiary study} & {\small 33 secondary studies} & {\small 2010} & {\small \cite{Kitchenham10}} & {\small Included} \\ [0.1pt] \hline 
		{\small Critical appraisal of SLRs in SE from the perspective of the research questions} & {\small 53 secondary studies} & {\small 2010} & {\small \cite{daSilva10}} & {\small Excluded\textsuperscript{(1)}} \\ [0.1pt] \hline 
		{\small Research synthesis in SE -- A tertiary study} & {\small 49 secondary studies} & {\small 2011} & {\small \cite{Cruzes11}} & {\small Included} \\ [0.1pt] \hline 
		{\small Six years of SLRs in SE -- An updated tertiary study} & {\small 67 secondary studies} & {\small 2011} & {\small \cite{Silva10}} & {\small Included} \\ [0.1pt] \hline 
		{\small Signs of Agile Trends in Global SE Research -- A Tertiary Study} & {\small 12 secondary studies} & {\small 2011} & {\small \cite{Hanssen11}} & {\small Included} \\ [0.1pt] \hline 
		{\small Systematic approach for identifying relevant studies in SE} & {\small 38 secondary studies} & {\small 2011} & {\small \cite{Zhang11}} & {\small Excluded\textsuperscript{(2)}} \\ [0.1pt] \hline 
		{\small SLRs in Distributed SE -- A Tertiary Study} & {\small 14 secondary studies} & {\small 2012} & {\small \cite{Marques12}} & {\small Included} \\ [0.1pt] \hline 
		{\small A tertiary study: experiences of conducting SLRs in SE} & {\small 116 secondary studies} & {\small 2013} & {\small \cite{Imtiaz13}} & {\small Included\textsuperscript{(3)}} \\ [0.1pt] \hline 
		{\small Risks and risk mitigation in global software development: A tertiary study} & {\small 34 secondary studies} & {\small 2014} & {\small \cite{Verner14}} & {\small Included} \\ [0.1pt] \hline \hline
		\multicolumn{5}{|p{15cm}|}{\small \textbf{Reasons for exclusions:}} \\ [0.1pt] 
		\multicolumn{5}{|p{15cm}|}{\textsuperscript{(1)} \small The authors analyzed 53 literature reviews that had been collected in two published tertiary studies (Kitchenham et al. 2009, 2010), both already included in our population.}  \\ 
		\multicolumn{5}{|p{15cm}|}{\textsuperscript{(2)} \small The authors conducted a search of SLRs published in SE, which extends the search reported in the tertiary study (Kitchenham et al., 2011, 20 SLRs) with the updated records by the end of 2008. This up-to-date SLR search identified 38 SLRs. The list of the updated studies were not available. However, the (Kitchenham et al., 2011)' study was extended by two other studies, already included in our population (Kitchenham et al., 2010 and da Silva et al., 2011).} \\ 
		\multicolumn{5}{|p{15cm}|}{\textsuperscript{(3)} \small The list of studies were not available. However, due to the number of secondary studies included (116) and the general theme about SE, we decided to request the list to the authors, who kindly sent it to us.
		} \\ [0.1pt] \hline 
	\end{tabular}
\end{table*}

39 documents resulted from the automatic search and of this total, 12 tertiary studies were found (see Table \ref{tbl:population1}). The range of secondary studies search was 2004--2017, a total of 10 tertiary studies were included, and 888 secondary studies were revealed. In summary, we added 18 tertiary studies (source 1: 8 + source 2: 10) and 1233 secondary studies (source 1: 345 + source 2: 888). 

\begin{table*} [!ht]
	\centering
	\caption{Secondary studies identified through an automatic search}
	\label{tbl:population1}
	\begin{tabular}{|p{6cm}|c|c|p{2cm}|}
		\hline
		\textbf{{\small Study}} & \textbf{{\small Total}} & \textbf{{\small Year}} & \textbf{{\small Ref.}} \\ [0.1pt] \hline 	\hline 	
		{\small SLRs in global software development: A tertiary study} & {\small 24 secondary studies} & {\small 2012} & {\small \cite{Verner12}} \\ [0.1pt] \hline
		{\small An SLR of SLR process research in SE} & {\small 68 secondary studies} & {\small 2013} & {\small \cite{Kitchenham13}} \\ [0.1pt] \hline 
		{\small SLRs in requirements engineering: A tertiary Study} & {\small  53 secondary studies} & {\small 2014} & {\small \cite{Bano14}} \\ [0.1pt] \hline 
		{\small Quality Assessment of SLRs in SE: A Tertiary Study} & {\small 127 secondary studies} & {\small 2015} & {\small \cite{Zhou15}} \\ [0.1pt] \hline 
		{\small A map of threats to validity of SLRs in SE} & {\small 316 secondary studies} & {\small 2016} & {\small \cite{Zhou16} \textsuperscript{(1)}} \\ [0.1pt] \hline 
		{\small Practical similarities and differences between SLRs and SMs: a tertiary study} & {\small 170 secondary studies} & {\small 2017} & {\small \cite{Napoleao17}} \\ [0.1pt] \hline 
		{\small Systematic Studies in Software Product Lines: A Tertiary Study} & {\small 60 secondary studies} & {\small 2017} & {\small \cite{Marimuthu17}} \\ [0.1pt] \hline
		{\small Reporting SLRs: Some lessons from a tertiary study} & {\small 37 secondary studies} & {\small 2017} & {\small \cite{Budgen18a}} \\ [0.1pt] \hline 
		{\small How do Secondary Studies in SE report Automated Searches?} & {\small 171 secondary studies} & {\small 2018} & {\small \cite{Singh18}} \\ [0.1pt] \hline 
		{\small A tertiary study on technical debt} & {\small 13 secondary studies} & {\small 2018} & {\small \cite{Rios18}} \\ [0.1pt] \hline 		
		{\small Identifying, categorizing and mitigating threats to validity in SE secondary studies} & {\small 165 secondary studies} & {\small 2018} & {\small \cite{Ampatzoglou18}} \\ [0.1pt] \hline 		
		{\small A tertiary study on model-based testing areas, tools and challenges} & {\small 10 secondary studies} & {\small 2018} & {\small \cite{Villalobos-Arias18}} \textsuperscript{(2)} \\ [0.1pt] \hline \hline
		\multicolumn{4}{|p{12cm}|}{\small \textbf{Reasons for exclusions:}} \\ [0.1pt] 
		\multicolumn{4}{|p{12cm}|}{\textsuperscript{(1)} \small The list of secondary studies was not available.} \\ [0.1pt] 
		\multicolumn{4}{|p{12cm}|}{\textsuperscript{(2)} \small Download of paper is not available.} \\ [0.1pt] \hline 
	\end{tabular}
\end{table*}

After identifying the secondary studies, we collected the main authors' names of this set of studies. Studies that did not provide the first author's email, we used the email of the corresponding author. A total of 555 authors were considered as not duplicated. These 555 authors formed our population and  they were contacted by email. 

\textbf{-- Phase 6:} Analyzing the data -- The answers were stored directly after they had been submitted by the respondents. A total of 11 respondents (17.19\%) wrote free comments and all these comments were also analyzed.

\subsection{Main Results}\label{results}

The survey was carried out between November 17$^{th}$ and December 7$^{th}$ 2018. 555 emails were sent, but 164 returned with shipping errors. We believe that this happened because many of the authors changed place of work, research group, etc., and consequently changed their email account. We received in total 64 responses from 391 survey emails sent successfully (16.37\%). 

The majority of the respondents, with 25 (39.10\%) are Faculty Members (Supervisors) followed by Graduate Students (Ph.D) with 21 (32.80\%); Graduate Students (Masters) with 9 (14.10\%) and Research Staff 6 (9.40\%). We received 3 unique responses of respondents who classified themselves as Fresh Ph.D. Graduates (1.60\%), Professors (1.60\%) and General Managers (1.60\%). None of the respondents are undergraduate students.

With the exception of only one respondent that mentioned ``Startups'', all other respondents cited ``Software Engineering'' as their research area. 26 (40.62\%) cited just ``Software Engineering'' as itself and the rest of 37 (57.81\%) mentioned Software Engineering subareas such as ``Empirical Software Engineering'', ``Requirements Engineering'', ``Software Architecture'', ``Software Testing'', ``Software Quality'', etc. In addition, more than half of the participants have more than 10 years of experience in the mentioned research area. Those with less time of experience mentioned 2--3 years. The average time of experience was 10.46 years. 

With respect to conducting secondary studies, 10 (15.62\%) participants have conducted 10 or more secondary studies, 7 (10.94\%) have conducted between 5--9, 24 (37.50\%) between 8--13 and the remainder of the 23 (35.94\%) respondents mentioned having conducted only one or two secondary studies. The total number of secondary studies conducted by the respondents were 274. Out of the 274 secondary studies, 185 (67.52\%) studies were published in conferences or journals. One respondent who mentioned conducting three secondary studies did not report the number of published studies.

\begin{figure*}[!h]
	\centering
	\includegraphics [width=0.9\textwidth]{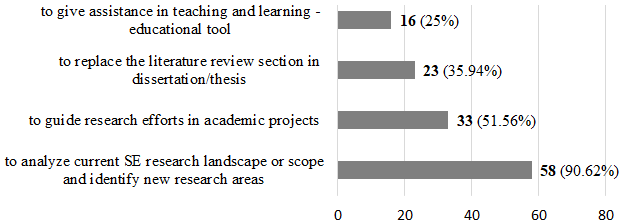}
	\caption{\textcolor{black}{Respondents’ perception about the academic purpose of the application of secondary studies in SE}}
	\label{PurposeSurvey}
\end{figure*}

Participants were asked to indicate what was the academic purpose of the application of secondary studies in SE. \textcolor{black}{The  vast  majority  (see  Figure  \ref{PurposeSurvey}) (58 -– 90.62\%) agreed that the purpose was to analyze current SE research landscape or scope and identify new research areas. Other purposes appeared in the sequence: to guide research efforts in academic projects (33 -- 51.56\%), to replace the literature review section in dissertation/thesis (23 -- 35.94\%), to give assistance in teaching and learning – educational tool (16 -- 25.00\%).} One respondent argued that he/she understands that ``\textit{nowadays the SLRs have been used for many different purposes, however he/she believes that the main purpose should be to analyze current SE research landscape or scope and identify new research areas as well as answering a research question (synthesizing knowledge)}''. Another respondent mentioned that in his/her university ``\textit{a secondary study is used to show that the work is unique due to an indirect requirement}''. Two others respondents emphasized ``\textit{the use of secondary studies is to identify the state of the art or establish the borderline of a research area of interest}''. 

\begin{figure*}[!ht]
	\centering
	\includegraphics [width=1.0\textwidth]{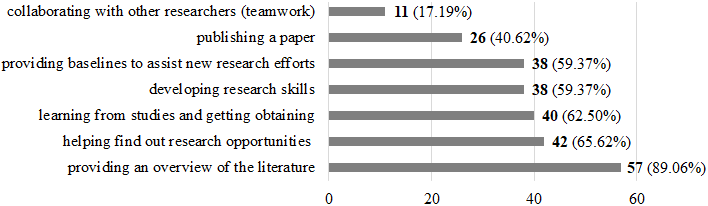}
	\caption{\textcolor{black}{Respondents’ perception about main benefits of conducting a secondary study during a research project}}
	\label{BenefitsSurvey}
\end{figure*}

\textcolor{black}{Regarding the  main benefits  of conducting a  secondary study during  a research project (see Figure \ref{BenefitsSurvey}), the biggest benefit highlighted is providing an  overview  of  the  literature  (57 -- 89.06\%)  followed  by  helping  find  out research opportunities (42 -- 65.62\%), learning from studies and getting obtaining (40 -– 62.50\%), developing research skills (38 -- 59.37\%) and providing baselines to assist new research efforts (38 -– 59.37\%).  One interesting point is that 26 (40.62\%) respondents agreed that publishing a paper as a main benefit of conducting a secondary study.  The benefit of collaborating with other researchers (teamwork) appeared as the lowest rate with 11 (17.19\%) responses.}

\begin{figure*}[!ht]
	\centering
	\includegraphics [width=1.0\textwidth]{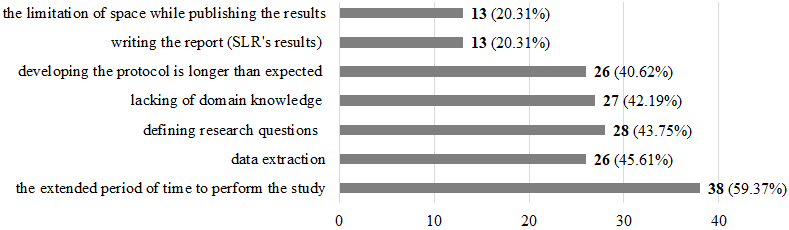}
	\caption{\textcolor{black}{Respondents’ perception about main problems of conducting a secondary study during a research project}}
	\label{ProblemsSurvey}
\end{figure*}

\textcolor{black}{The  main  problem  of  conducting  a  secondary  study  during  a  research project according to the respondents is the extended period of time to perform the study (38 -- 59.37\%) (see Figure \ref{ProblemsSurvey}). Other  problems  include:  data  extraction  (26  --  45.61\%);  defining  research questions (28 -- 43.75\%);  lacking of domain knowledge (27 -– 42.19\%);  and developing  the  protocol  is  longer  than  expected  (26  -–  40.62\%). Other less significant problems cited are related to the last phase of SLR: writing the report (SLR's results) and the limitation of space while publishing the results, both with 13 responses (20.31\%).} Particularly in the comments, two respondents mentioned problems related to the search process in digital libraries due to the difficulty of performing automatic searches requiring different keywords combination and unavailability of some publications. \textcolor{black}{Other less significant problems cited are related to the last phase of SLR: writing the report (SLR's results) and the space limitations when publishing the results, both with 13 responses (20.31\%).} 

\begin{figure*}[!ht]
	\centering
	\includegraphics [width=1\textwidth]{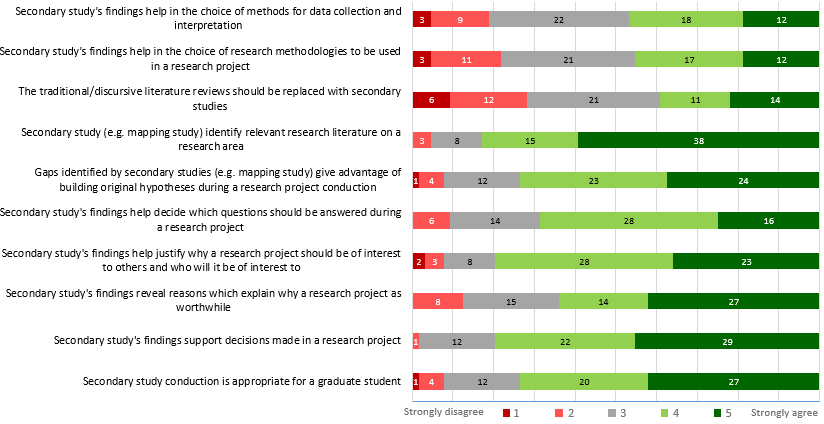}
	\caption{Questions and responses regarding the respondents' perception about the use of secondary studies}
	\label{Results}
\end{figure*}

The third section of the questionnaire presents respondents' perception about the use of secondary studies. The responses are based on Likert Scale method (1 to 5, where 1 means ``Strongly disagree'' and 5 ``Strongly agree''). The responses of each question are illustrated in Figure \ref{Results}. In summary, most of the questions presented in this section of the questionnaire had positive responses (4 and 5 scales).

Reflecting on the questions: ``The traditional/discursive literature reviews should be replaced with secondary studies"; ``Secondary studies findings help in the choice of research methodologies to be used in a research project"; and ``Secondary studies findings help in the choice of methods for data collection and interpretation" (the first three questions in Figure \ref{Results}), the biggest concentration of answers are in scale 3, which means a neutral opinion about them.  The question with more positive answers is ``Secondary studies (e.g. mapping study) identify relevant research literature on a research area'' with 15 (23.44\%) scale 4 (agree) answers and 38 (59.37\%) answers scale 5 (strongly agree) out 64, totalling 53 (82.81\%) answers. 

Considering the free comments, 11 respondents wrote answers. Overall, all respondents in some way highlighted the importance of conducting secondary studies. One of them explained in details how he/she used secondary study in his/her Ph.D. thesis saying: ``First I conducted a mapping study, second I detailed this mapping study. These were very important to define the theme and research questions to my thesis. Finally, I performed a systematic review to discuss the results (results from case studies and results from third parties)''.  Other two researchers mentioned that secondary studies are a valuable tool for M.Sc. and Ph.D. students stating ``I think every M.Sc./Ph.D. student must perform mapping study before proposing an idea''.

They emphasize some points that have to be considered with attention by secondary studies practitioners: -- Applicability of the study: ``I think that it is important to consider the applicability of the secondary studies. What if an area is new? Then there are no benefits of conducting a secondary study. What if it is very large? What if numerous literature reviews exist?''; ``It is necessary to carefully evaluate when conducting a secondary study is actually appropriate''; ``Secondary studies must be conducted properly and the purpose of performing a secondary study must be well-established. Otherwise, a lot of effort can be spent to produce results that are not meaningful to the research project''; -- Avoid bias: ``Nowadays, it is common to see one student performing a secondary study just with his/her advisor that ``review'' only part of the findings. Therefore, for me, the biggest challenge is: How to avoid the researcher bias?''; ``Depending on the research topic, other stakeholders such as practitioners' input should be considered in addition to the research gaps found in the secondary studies''. 

The main contribution of this survey was the answers by SE researchers including faculty members (supervisors), Ph.D./M.Sc. graduates, among others and their opinions offered us an opportunity to better understand how secondary studies have been inserted and applied in the academic context. Based on the data provided by the respondents, the biggest contribution that secondary studies can bring is to provide an overview of the literature as well as to identify relevant research literature on a research area enabling to find reasons to explain why a research project should be provided and/or to support decisions made in a research project. One point worth highlighting is that the respondents agree or strongly agree that conducting secondary studies is appropriate for a graduate student, mainly Ph.D. students, due to the contribution that they can bring to the investigation, for example, the identification of gaps for building original hypotheses. 

\subsection{Discussions}\label{discussions}

The results provide interesting insights for conducting secondary studies including application for academic purpose, practical problems and benefits of and perceptions about their use.

\begin{figure*}[!ht]
	\centering
	\includegraphics [width=1\textwidth]{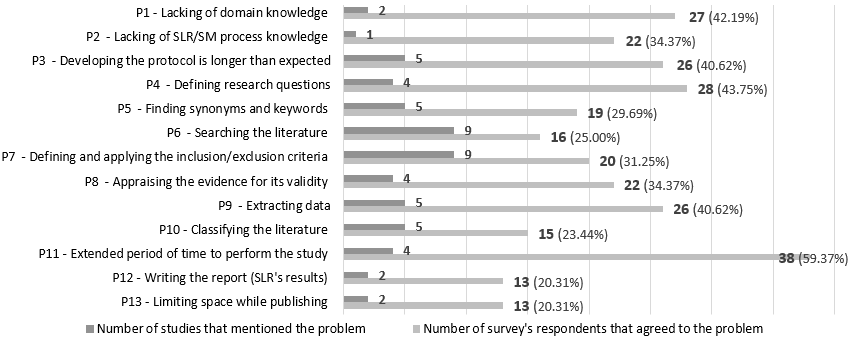}
	\caption{Survey respondents' agreement x number of studies that mention the problem}
	\label{Problems}
\end{figure*}

\begin{figure*}[!ht]
	\centering
	\includegraphics [width=1\textwidth]{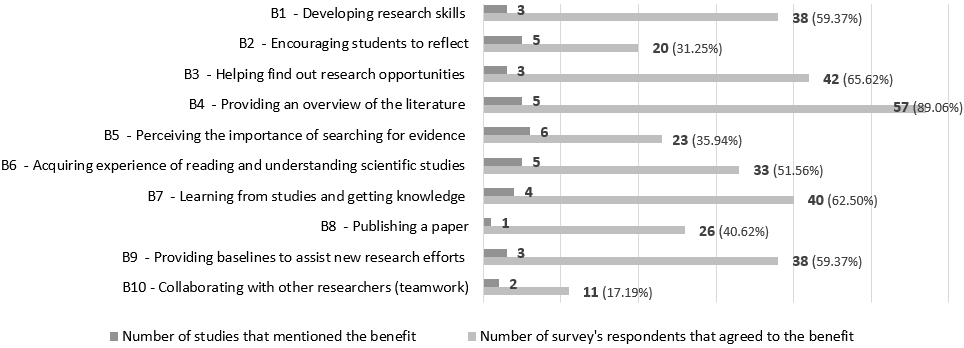}
	\caption{Survey respondents' agreement x number of studies that mention the benefit}
	\label{Benefits}
\end{figure*}

The relationship between survey respondents' appointments and the number of studies (SMS) that mention a particular problem/benefit is illustrated in Figures \ref{Problems} and \ref{Benefits}.

Comparing the results presented in Figure \ref{Problems}, it is possible to observe that all problems listed in the studies resulted from the SMS were pointed out by the survey respondents. The problems ``P5 -- Searching the literature'' and ``P6 -- Defining and applying the inclusion/exclusion criteria'' had the highest appearance in the studies (9 studies), however, the number of key points of both problems is low (P5 = 25.00\% and P6 = 31.25\%). The same behavior can be observed in  ``P4 -- Defining the research questions'' with 28 (42.19\%) key points and 4 studies that mention it. 

``P1 -- Lacking of domain knowledge'' was the third most problem pointed out by survey respondents, however it was mentioned by 2 studies only. The problem ``P2 -- Lacking of SLR/SMS process knowledge'' was mentioned by a unique study but 22 (34.37\%) respondents considered it a problem. The problems  ``P12 -- Writing the report (SLR's results)'' and ``P13 -- Limiting space while publishing'' had a more balanced results with 2 mentions in both key points by the survey respondents and studies.

According to the survey respondents the biggest problem faced is ``P11 -- Extended period of time to perform the study'' with 33 (59.37\%) key points, and, this problem was mentioned by 4 studies. While a secondary study provides a methodical and reliable way of conducting the literature review, the process is rather intensive when compared to a traditional review. On one hand, secondary studies require, e.g., the definition of a protocol in the planning phase, while a traditional review does not necessarily require a protocol definition. Moreover, activities, such as the definition of research questions, the creation of search strings, and the selection of study repositories, are non-trivial. Nevertheless, reviews based on a systematic approach result in a rigorous process, that controls, for example, the following activities: searching for studies on the Ph.D. research topic; reading and summarizing the main points from relevant studies; synthesizing the key ideas, theories and concepts; discussing and evaluating these ideas, theories and concepts. On the other hand, during a non-systematic review there is no control on how the tasks are run. 

Despite the difficulties, one of the reasons why secondary studies should be conducted by Ph.D. students compared to informal reviews is their advantages, which include the identification of gaps in current research, which may suggest new research themes and provide a suitable way to position such themes in the context of existing research (state-of-the-art). Furthermore, conducting secondary studies develops important skills, such as, ability to search, identify, understand, critically appraise, and integrate data, for Ph.D. students. 

We identified 21 experiences of Ph.D. students who have conducted a secondary study as part of their literature review for their Ph.D. thesis. The students argued that SLRs are not quick to conduct and, depending on the extent of the relevant literature, can take months to complete. As a main advantage, they emphasize that SLRs are useful in providing evidence on a research topic. 

The same comparison regarding the number of studies from SMS performed and survey respondents' appointments was considered to the benefits, as shown in Figure \ref{Benefits}.

All benefits mentioned in the studies (SMS) were pointed out by the survey respondents (see Figure \ref{Benefits}). The benefit ``B4 -- Providing an overview of the literature'' was the most pointed out benefit by the survey respondents with 57 (89.06\%) and also is one of the most mentioned benefit in the studies (5 studies). Indeed, secondary studies help to find out research opportunities as well as to provide baseline to assist new research efforts due to the overview of the state of art that secondary studies bring, even an SMS that is more general.

Benefit ``B5 -- Perceiving the importance of searching for evidence''  was the most mentioned benefit in the studies (6 studies), however it is one of the lowest pointed out by the surveys respondents (third less mentioned with 23 (35.94\%) mentions). The benefit ``B10 -- Collaborating with other researchers (teamwork)'' had the lowest number of key points by survey respondents and also a low number of mentions in studies (2 mentions only). The benefit ``B8  -- Publishing a paper'' was mentioned by only one study in SMS, however for the survey  respondents it is an important point since this benefit was pointed out by 26 (40.62\%) of the survey respondents. 

In summary, we can say that the top three benefits (B3, B4 and B7) pointed out by the survey respondents agree with the benefits pointed out in the literature, since that all three benefits had appeared in 3 or more studies from the secondary studies we performed. 

It is common knowledge that the incentive to train new researchers begins during undergraduate courses. Since research is an essential component of doctoral and master's degree programs, having an undergraduate research experience is a valuable opportunity. One possible way to improve the research skills of undergraduate students would be to design disciplines covering SLR. Scientific activity enriches the student's curriculum and improves their academic training. This practice also enables the integration of teaching and research activities in the university and the training of more critical professionals. Encouraging undergraduate students to conduct an SLR can positively influence graduates (M.Sc./Ph.D.) learning.

We had positive experiences from undergraduate students who conducted secondary and tertiary studies to their completion of course work. Examples are \cite{Napoleao17, Mariano15}. In common these two students claimed that they gained educational benefits including, among others: $\bullet$ improving writing skill; $\bullet$ improving English skill; $\bullet$ contacting with specific research topics not covered at the undergraduate level (increased motivation), among others. Both undergraduate students started a postgraduate course. One contributing factor was that a secondary/tertiary can identify current research gaps more precisely and, as a consequence, help define future research initiatives. 

Some of threats to the validity are:

\begin{itemize}
	\item Missed related works: Some of the relevant studies related to the topic of this research could be missed. To mitigate this issue we conducted a two-stage search strategy: automatic search and snowballing. The automatic search resulted in three studies and the snowballing application provided 25 relevant studies. We believe that most of the relevant studies were covered by our search strategy. Moreover, our seed set for snowballing was carefully created considering the studies identified as relevant in automatic search;
	
	\item Construct validity: Survey options do not cover all possibilities. However, both the direct and those Likert Scale questions emerged from the SMS performed. We inserted some open questions and at the end of the survey a free space for final comments where the respondent could report any additional information without limitation;
	
	\item External validity: The survey was conducted between November 17$^{th}$ and December 7$^{th}$ 2018 including the first or the corresponding author of published secondary studies. Thus, we assumed that the respondents were researchers with some experience in conducting secondary studies in SE thus they were considered a suitable survey population. The response rate was 16.37\% among authors successfully contacted via email. Therefore, we consider the result satisfactory. Another limitation is that not all recipients received the survey email successfully. In order to mitigate this threat of sending the email from a tool and these emails be blocked by anti-spam tools or firewalls, all emails were sent manually.
\end{itemize}

\section{Conclusions}\label{conclusions}

\textcolor{black}{We administered a survey about the use of secondary studies in academic projects and the findings from this study are expected to contribute to the existing knowledge about the use and contribution of secondary studies in the academy.} In summary, we concluded that secondary studies outcomes can be useful for M.Sc./Ph.D. students. Main reasons are: 

\begin{enumerate}	
	\item Providing a structure for undertaking and writing broad literature reviews; 
	
	\item Demonstrating gaps in the literature, which can then help to improve the design and justify the research or find out further opportunities worth investigating;
	
	\item Learning more about a certain subject of interest;
	
	\item Identifying relevant research literature on a research area; and findings can support decisions made in a research project; 
	
	\item Gaining knowledge and practical research skills; and
	
	\item Building connections with researchers,  M.Sc./Ph.D. students, and other undergraduates who share similar research interests. 
	
\end{enumerate}

The acquisition of these skills and benefits increases students' motivation to pursue their research project and prepares them to both, academic or industrial careers. Although there are publications defining how to conduct secondary studies \cite{Kitchenham07,Petersen08, Petersen15}, there is still a need for studies addressing the main problems faced by secondary study practitioners mainly in order to provide better support to  M.Sc./Ph.D. students. Among these problems we can emphasize the effort/time spent to conduct such studies, which is greater when compared to a traditional review. 

In conclusion, secondary studies have a number of advantages in an academic context and therefore  M.Sc./Ph.D. students should consider doing a secondary study for their research project, of course, under the supervision of more experienced researchers.

Other contributions of this paper can be highlighted as follows: 

\begin{enumerate}
	\item We found a more recent list of secondary studies in SE, including 1233 secondary studies; and
	\item A common problem in conducting survey is sampling, in particular, the ability to obtain a sufficiently large sample \cite{Amir18} (population lists -- corpus). We believe that we have defined the actual sampling of researchers that perform secondary studies in SE.
\end{enumerate}	

\textcolor{black}{One concern is how to adjust EBSE teaching in SE courses. One point that should be considered is to use evidence to solve a relevant SE problem or need. The focus of the course should not only be to teach EBSE, research skills, but also how to integrate evidence and practical experience. As a future work we intend to design a course for graduate students (M.Sc. and Ph.D.) covering SLR and SMS. We believe that this course can improve their research skills and help them with their problems and challenges in conducting secondary studies. As a consequence, influencing positively learning, improving their academic training.}

\begin{center} ACKNOWLEDGMENT \end{center} 
			
The authors thank the financial support received from CNPq (projects 401033/2016-3 and 432247/2018-1), Brazil. \\




\bibliographystyle{elsarticle-num-names} 
\bibliography{referencias}





\end{document}